\documentclass{article}
\usepackage{amsmath,amssymb,theorem,mathpazo,color,url}
\allowdisplaybreaks
\newtheorem{thm}{Theorem}[section]
\newtheorem{cor}[thm]{Corollary}
\newtheorem{lem}[thm]{Lemma}
\newtheorem{prop}[thm]{Proposition}
\newtheorem{defn}[thm]{Definition}
\newtheorem{rem}[thm]{Remark}
\newtheorem{ass}[thm]{Assumption}

\def\proof {\noindent{\it Proof.}$\quad$}
\def\fin   {\hfill{$\Box$}\vspace{3mm}}
\def\l     {\left}
\def\r     {\right}
\def\<     {\langle}
\def\>     {\rangle}

\def\supp  {\mathop{\mathrm{supp}}\nolimits}
\def\calB  {{\cal B}}

\def\calF  {{\cal F}}

\def\bbD   {{\mathbb D}}
\def\bbE   {{\mathbb E}}
\def\bbF   {{\mathbb F}}

\def\bbN   {{\mathbb N}}
\def\bbP   {{\mathbb P}}

\def\bbR   {{\mathbb R}}
\def\ve    {\varepsilon}
\def\vp    {\varphi}

\def\tP    {\bbP^*}
\def\tN    {\widetilde{N}}

\begin{document}
\title{Local risk-minimization for Barndorff-Nielsen and Shephard models
       with volatility risk premium}
\author{Takuji Arai\footnote{
        Department of Economics, Keio University, e-mail:arai@econ.keio.ac.jp}
}

\maketitle

\begin{abstract}
We derive representations of local risk-minimization of call and put options
for Barndorff-Nielsen and Shephard models:
jump type stochastic volatility models whose squared volatility process
is given by a non-Gaussian Ornstein-Uhlenbeck process.
The general form of Barndorff-Nielsen and Shephard models includes
two parameters: volatility risk premium $\beta$ and leverage effect $\rho$.
Arai and Suzuki \cite{AS-BNS} dealt with the same problem under
constraint $\beta=-\frac{1}{2}$.
In this paper, we relax the restriction on $\beta$;
and restrict $\rho$ to $0$ instead.
We introduce a Malliavin calculus under the minimal martingale measure
to solve the problem.
\end{abstract}


\setcounter{equation}{0}
\section{Introduction}
Local risk-minimization (LRM, for short) for Barndorff-Nielsen and
Shephard models (BNS model, for short) is discussed.
Here LRM is a very well-known quadratic hedging method
of contingent claims for incomplete financial markets.
On the other hand, BNS models are stochastic volatility models
suggested by Barndorff-Nielsen and Shephard \cite{BNS1}, \cite{BNS2}.
It is known that some stylized facts of financial time series are captured
by BNS models.
The square volatility process $\sigma^2$ of a BNS model is given as an
Ornstein-Uhlenbeck process driven by a subordinator without drift, that is,
a nondecreasing pure jump L\'evy process.
Thus, $\sigma^2$ is a jump process given as a solution to the following
stochastic differential equation (SDE, for short):
\[
d\sigma_t^2=-\lambda\sigma_t^2dt+dH_{\lambda t}, \ \ \ \sigma_0^2>0,
\]
where $\lambda>0$, $H$ is a subordinator without drift.
Now, we denote by $S$ the underlying asset price process.
The general form of $S$ is given by
\[
S_t=S_0\exp\l\{\int_0^t\l(\mu+\beta\sigma_s^2\r)ds+\int_0^t\sigma_sdW_s
    +\rho H_{\lambda t}\r\},
\]
where $S_0>0$, $\mu$, $\beta\in\bbR$, $\rho\leq0$,
$W$ is a $1$-dimensional Brownian motion.
The last term $\rho H_{\lambda t}$ accounts for the leverage effect;
and $\beta\sigma_s^2$ is called the volatility risk premium,
which is considered as the compensation required by investors holding
volatile assets.
From the view of (\ref{SDE}) below, the volatility risk premium is vanished
when $\beta=-\frac{1}{2}$.
So that, $\beta$ would take a value greater than or equal to $-\frac{1}{2}$.
For more details on BNS models, see Cont and Tankov \cite{CT} and
Schoutens \cite{Scho}.

Our purpose is to obtain representations of LRM of call and put options
for BNS models under constraint $\rho=0$ and no constraint on $\beta$.
On the other hand, Arai and Suzuki \cite{AS-BNS} studied the same problem
under constraint $\beta=-\frac{1}{2}$ and no constraint on $\rho$.
That is, they dealt with the case where volatility risk premium is not
taken into account.
To the contrary, we will treat BNS models with volatility risk premium.
In other words, we relax the restriction on $\beta$.
Instead, we restrict $\rho$ to $0$, which induces the continuity of $S$.
Then, $S$ is written as
\begin{equation}
\label{eq-S}
S_t=S_0\exp\l\{\int_0^t\l(\mu+\beta\sigma_s^2\r)ds+\int_0^t\sigma_sdW_s\r\}.
\end{equation}
Actually, the continuity of $S$ makes the problem easy to deal with.
To calculate LRM, we need to consider the minimal martingale measure
(MMM, for short).
When $S$ is continuous, the subordinator $H$ remains a L\'evy process
even under the MMM.
On the other hand, the generalization of $\beta$ makes the problem complicated.
When $\beta=-\frac{1}{2}$, the density process $Z$ of the MMM is given as a
solution to an SDE with the Lipschitz continuity.
Thus, as shown in \cite{AS-BNS}, $Z$ has the Malliavin differentiability,
which played a vital role in \cite{AS-BNS}.
However, this property is not generalized to the case of
$\beta\neq-\frac{1}{2}$.
Hence, we need to take a different approach from \cite{AS-BNS}.
In order to overcome this difficulty, making the best of the fact
that the L\'evy property of $H$ is preserved,
we innovate a Malliavin calculus under the MMM.
As a result, we can calculate LRM without attention to the property of $Z$.

To our best knowledge, except for \cite{AS-BNS},
there is only one preceding research on LRM for BNS models:
Wang, Qian and Wang \cite{WQW}.
Besides they treated the problem under the same parameter restrictions as ours,
although they did not use Malliavin calculus.
However, their discussion seems to be inaccurate mathematically.

Outline of this paper is as follows.
A precise model description and standing assumptions are given in Section 2.
In Subsections 2.1 -2.3, we define LRM, the MMM and a Malliavin derivative,
respectively.
Our main results are provided in Section 3; and
conclusions will be given in Section 4.

\setcounter{equation}{0}
\section{Preliminaries}
We consider a financial market model in which only one risky asset and
one riskless asset are tradable.
For simplicity, we assume that the interest rate is given by $0$.
Let $T>0$ be the finite time horizon.
The fluctuation of the risky asset is described as a process $S$
given by (\ref{eq-S}).
We consider a complete probability space $(\Omega, \calF, \bbP)$ with
a filtration $\bbF=\{\calF_t\}_{t\in[0,T]}$ as the underlying space.
Suppose that $\bbF$ is generated by $W_t$ and $H_{\lambda t}$;
and satisfies the usual condition, that is, $\bbF$ is right continuous, and
$\calF_0$ contains all null sets of $\bbP$.
The asset price process $S$ given in (\ref{eq-S})
is a solution to the following SDE:
\begin{equation}
\label{SDE}
dS_t=S_{t-}\l\{\mu dt+\l(\beta+\frac{1}{2}\r)\sigma^2_tdt+\sigma_tdW_t\r\}.
\end{equation}
Denoting $A_t:=\int_0^tS_{s-}\l[\mu+\l(\beta+\frac{1}{2}\r)\sigma^2_s\r]ds$
and $M_t:=S_t-S_0-A_t$, we have $S_t=S_0+M_t+A_t$,
which is the canonical decomposition of $S$.
Further, we denote $L_t:=\log(S_t/S_0)$ for $t\in[0,T]$, that is,
\[
L_t=\mu t+\beta\int_0^t\sigma_s^2ds+\int_0^t\sigma_sdW_s.
\]
Defining $J_t:=H_{\lambda t}$, we denote by $N$ the Poisson random measure
of $J$, that is, we have $J_t=\int_0^\infty xN([0,t],dx)$.
Denoting by $\nu$ the L\'evy measure of $J$,
we have that $\tN(dt,dx):=N(dt,dx)-\nu(dx)dt$ is
the compensated Poisson random measure.
Remark that $N$ and $\nu$ are defined on $[0,T]\times(0,\infty)$
and $(0,\infty)$, respectively; and $\nu(dx)=\lambda\nu^H(dx)$,
where $\nu^H$ is the L\'evy measure of $H$.
Moreover, Proposition 3.10 of \cite{CT} implies
\begin{equation}
\label{eq-nu-cond}
\int_0^\infty(x\wedge1)\nu(dx)<\infty.
\end{equation}

We need to impose the following standing assumptions on $\nu$
as in \cite{AS-BNS}.
As stated in Remark \ref{rem1} below, the standing assumptions do not exclude
representative examples of BNS models, although parameters are restricted.

\begin{ass}
\label{ass1}
\begin{description}
\item[(A1)] 
The L\'evy measure $\nu$ is absolutely continuous with respect to the
Lebesgue measure on $(0,\infty)$.
\item[(A2)] 
There exists a $\kappa>0$ such that
\begin{itemize}
\item
$\kappa>\l[2\l(\beta+\frac{1}{2}\r)^++1\r]\calB(T)$,
\item
$\kappa\geq\l(\beta+\frac{1}{2}\r)^2\calB(T)$, and
\item
$\int_1^\infty e^{2\kappa x}\nu(dx)<\infty$,
\end{itemize}
where $\calB(t):=\int_0^te^{-\lambda s}ds=\frac{1-e^{-\lambda t}}{\lambda}$
for $t\in[0,T]$.
\end{description}
\end{ass}

\begin{rem}
\label{rem1}
\begin{enumerate}
\item
When $\beta=-\frac{1}{2}$, (A2) is equivalent to the existence of
$\ve>0$ such that $\int_0^\infty e^{(2+\ve)\calB(T)x}\nu(dx)<\infty$.
In \cite{AS-BNS} dealing with the case of $\beta=-\frac{1}{2}$,
$\int_0^\infty e^{2\calB(T)x}\nu(dx)<\infty$ is assumed
in their Assumption 2.2, which is almost same as
the above (A2) for $\beta=-\frac{1}{2}$.
\item
We do not need to assume conditions corresponding to the second condition
of Assumption 2.2 in \cite{AS-BNS}, which ensures the positivity of the density
of the MMM defined below,
since the MMM becomes a probability measure automatically in our setting.
\item
Condition (A2) ensures $\int_0^\infty x^2\nu(dx)<\infty$,
which means $\bbE[J_T^2]<\infty$.
In addition, we have $\bbE[e^{2\kappa J_T}]<\infty$
by Proposition 3.14 of \cite{CT}.
\item
Condition (A1) guarantees Assumption Z1 in Nocolato and Venardos \cite{NV},
which we need in the proof of Lemma \ref{lem7} below.
\item
Assumption \ref{ass1} does not exclude two representative examples of
$\sigma^2$, ``IG-OU" and ``Gamma-OU".
``IG-OU" is the case where $\nu^H$ is given as
\[
\nu^H(dx)=\frac{a}{2\sqrt{2\pi}}x^{-\frac{3}{2}}(1+b^2x)
          e^{-\frac{1}{2}b^2x}{\bf 1}_{(0,\infty)}(x)dx,
\]
where $a>0$ and $b>0$.
The invariant distribution of $\sigma^2$ follows an inverse-Gaussian
distribution with $a>0$ and $b>0$.
Then $\sigma^2$ is called an IG-OU process.
If
\[
\frac{b^2}{2}> 2\l\{\l[2\l(\beta+\frac{1}{2}\r)^++1\r]
               \vee\l(\beta+\frac{1}{2}\r)^2\r\}\calB(T),
\]
then Assumption \ref{ass1} is satisfied.
Next, ``Gamma-OU" is the case where the invariant distribution of $\sigma^2$
is given by a Gamma distribution with $a>0$ and $b>0$.
In this case, $\nu^H$ is described as
\[
\nu^H(dx)=abe^{-bx}{\bf 1}_{(0,\infty)}(x)dx.
\]
As well as the IG-OU case, Assumption \ref{ass1} is satisfied if
\[
b> 2\l\{\l[2\l(\beta+\frac{1}{2}\r)^++1\r]
   \vee\l(\beta+\frac{1}{2}\r)^2\r\}\calB(T).
\]
For more details on this topic, see also \cite{NV} and \cite{Scho}.
\end{enumerate}
\end{rem}

\subsection{Local risk-minimization}
In this subsection, we define LRM.
To this end, we define the SC condition firstly;
and show that $S$ satisfies it under Assumption \ref{ass1}.
$S$ is said to satisfy the SC condition,
if the following three conditions hold:
\begin{enumerate}
\item[(a)]
$\l\|[M]_T^{1/2}+\int_0^T|dA_s|\r\|_{L^2(\bbP)}<\infty$.
\item[(b)]
Defining a process $\Lambda_t
:=\frac{1}{S_{t-}}\frac{\mu+\l(\beta+\frac{1}{2}\r)\sigma^2_t}{\sigma_t^2}$,
we have $A=\int\Lambda d\langle M\rangle$.
\item[(c)]
The mean-variance trade-off process
$K_t:=\int_0^t\Lambda^2_sd\langle M\rangle_s$ is finite, that is,
$K_T$ is finite $\bbP$-a.s.
\end{enumerate}

\begin{prop}
\label{prop-SC}
$S$ satisfies the SC condition under Assumption \ref{ass1}.
\end{prop}

\proof
It suffices to show item (a) only.
Note that we have
\begin{align*}
\lefteqn{\l\|[M]_T^{1/2}+\int_0^T|dA_t|\r\|^2_{L^2(\bbP)}} \\
&\leq 2\bbE\l[[M]_T+\l(\int_0^T|dA_t|\r)^2\r] \\
&\leq 2\bbE\l[\int_0^TS_{t-}^2\sigma^2_tdt+\l(\int_0^TS_{t-}\l|\mu
      +\l(\beta+\frac{1}{2}\r)\sigma^2_t\r|dt\r)^2\r] \\
&\leq 2\bbE\l[\sup_{0\leq s\leq T}S_s^2\l\{\int_0^T\sigma^2_tdt
      +\l(|\mu|T+\l|\beta+\frac{1}{2}\r|\int_0^T\sigma^2_tdt\r)^2\r\}\r].
\end{align*}
If $\sup_{0\leq s\leq T}S_s\in L^{2a}(\bbP)$ holds
for a sufficiently small $a>1$,
item (a) holds by the H\"older inequality and Lemma \ref{lem3} below.

Now, we take an $a>1$ such that
\begin{equation}
\label{eq-prop-SC-1}
\l\{2\l(a\beta+\frac{a^2}{2}\r)^++a^2\r\}\calB(T)<\kappa.
\end{equation}
Note that we can find such an $a>1$ from the view of (A2)
in Assumption \ref{ass1}.
We shall see $\sup_{0\leq s\leq T}S_s\in L^{2a}(\bbP)$.
Since we have
\begin{align}
\int_0^t\sigma_s^2ds
&=    \sigma_0^2\int_0^te^{-\lambda s}ds
      +\int_0^t\int_0^se^{-\lambda(s-u)}dJ_uds \nonumber \\
&=    \sigma_0^2\calB(t)+\int_0^t\int_u^te^{-\lambda(s-u)}dsdJ_u
      =\sigma_0^2\calB(t)+\int_0^t\calB(t-u)dJ_u \nonumber \\
&\leq \sigma_0^2\calB(t)+\calB(t)J_t
      \leq \sigma_0^2\calB(T)+\calB(T)J_t
\label{eq-prop-SC-2}
\end{align}
for any $t\in[0,T]$, we obtain
\begin{align*}
e^{aL_t}
&=    \exp\bigg\{a\mu t+a\beta\int_0^t\sigma^2_sds+a\int_0^t\sigma_sdW_s\bigg\}
      \\
&=    \exp\bigg\{a\mu t-\frac{a^2}{2}\int_0^t\sigma^2_sds+a\int_0^t\sigma_sdW_s
      +\l(a\beta+\frac{a^2}{2}\r)\int_0^t\sigma^2_sds\bigg\} \\
&\leq C\exp\bigg\{-\frac{a^2}{2}\int_0^t\sigma^2_sds+a\int_0^t\sigma_sdW_s \\
&     \hspace{7mm}+\int_0^t\int_0^\infty bx\tN(ds,dx)
      +\int_0^t\int_0^\infty[bx+1-e^{bx}]\nu(dx)ds\bigg\} \\
&=:   CY^{a,b}_t,
\end{align*}
where $b:=\l(a\beta+\frac{a^2}{2}\r)^+\calB(T)$, and
$C:=\exp\{a|\mu|T+b\sigma^2_0+\int_0^T\int_0^\infty(e^{bx}-1)\nu(dx)dt\}$.
Taking into account of (\ref{eq-prop-SC-1}) and (A2) in Assumption \ref{ass1},
Lemma \ref{lem4} below yields that $Y^{a,b}$ is a square integrable martingale.
Thus, Doob's inequality yields
\begin{align*}
\bbE\l[\sup_{0\leq s\leq T}S_s^{2a}\r]
&=    \bbE\l[S_0^{2a}\sup_{0\leq s\leq T}e^{2aL_s}\r]
      \leq S_0^{2a}C^2\bbE\l[\sup_{0\leq s\leq T}(Y^{a,b}_s)^2\r] \\
&\leq 4S_0^{2a}C^2\bbE[(Y^{a,b}_T)^2]
      <    \infty.
\end{align*}
\fin

\begin{lem}
\label{lem3}
$\int_0^T\sigma^2_tdt\in L^n(\bbP)$ for any $n\geq1$.
\end{lem}

\proof
From the view of (\ref{eq-prop-SC-2}),
it suffices to show $J_T\in L^n(\bbP)$ for any $n\geq1$.
By Remark \ref{rem1}, we have $\bbE[\exp\{2\kappa J_T\}]<\infty$,
from which $J_T\in L^n(\bbP)$ follows for any $n\geq1$.
\fin

\begin{lem}
\label{lem4}
For $a\in\bbR$ and $b\geq0$, we denote
\begin{align*}
Y^{a,b}_t
&:= \exp\bigg\{-\frac{a^2}{2}\int_0^t\sigma^2_sds+a\int_0^t\sigma_sdW_s
    +\int_0^t\int_0^\infty bx\tN(ds,dx) \\
&   \hspace{7mm}+\int_0^t\int_0^\infty[bx+1-e^{bx}]\nu(dx)ds\bigg\}.
\end{align*}
\begin{enumerate}
\item 
If $a$ and $b$ satisfy
\begin{equation}
\label{eq-lem4-1}
\int_1^\infty\exp\l\{\l(2b+\frac{a^2}{2}\calB(T)\r)x\r\}\nu(dx)<\infty,
\end{equation}
then the process $Y^{a,b}$ is a martingale.
\item
When we strengthen (\ref{eq-lem4-1}) to
\begin{equation}
\label{eq-lem4-2}
\int_1^\infty\exp\{(4b+2a^2\calB(T))x\}\nu(dx)<\infty,
\end{equation}
$Y^{a,b}$ is a square integrable martingale.
\end{enumerate}
\end{lem}

\proof
{\it 1.}
From the view of Theorem 1.4 of Ishikawa \cite{I}, we need only to show that \\
(1) $\int_0^\infty[b^2x^2+(1-e^{bx})^2]\nu(dx)<\infty$, \\
(2) $\int_0^\infty[e^{bx}\cdot bx+1-e^{bx}]\nu(dx)<\infty$, and \\
(3) $\bbE\l[\exp\l\{\frac{a^2}{2}\int_0^T\sigma^2_tdt\r\}\r]<\infty$. \\
By (\ref{eq-nu-cond}) and (\ref{eq-lem4-1}),
conditions (1) and (2) are satisfied.
Next, (\ref{eq-lem4-1}) and Proposition 3.14 in \cite{CT} imply
$\bbE\l[\exp\l\{\frac{a^2}{2}\calB(T)J_T\r\}\r]<\infty$,
from which condition (3) follows by (\ref{eq-prop-SC-2}).

{\it 2.}
Denoting $\gamma:=2b+a^2\calB(T)$, we have
\begin{align*}
(Y^{a,b}_T)^2
&=    \exp\bigg\{-a^2\int_0^T\sigma_s^2ds+2a\int_0^T\sigma_tdW_t \\
&     \hspace{7mm}+\int_0^T\int_0^\infty2bx\tN(dx,dt)
      +\int_0^T\int_0^\infty2[bx+1-e^{bx}]\nu(dx)dt\bigg\} \\
&\leq \exp\bigg\{-2a^2\int_0^T\sigma_s^2ds+2a\int_0^T\sigma_tdW_t
      +a^2\sigma_0^2\calB(T)+a^2\calB(T)J_T \\
&     \hspace{7mm}+\int_0^T\int_0^\infty2bx\tN(dx,dt)
      +\int_0^T\int_0^\infty2[bx+1-e^{bx}]\nu(dx)dt\bigg\} \\
&=    \exp\bigg\{-2a^2\int_0^T\sigma_s^2ds+2a\int_0^T\sigma_tdW_t
      +\int_0^T\int_0^\infty\gamma x\tN(dx,dt) \\
&     \hspace{7mm}+\int_0^T\int_0^\infty\l[\gamma x+2-2e^{bx}\r]
      \nu(dx)dt+a^2\sigma_0^2\calB(T)\bigg\} \\
&=    \exp\bigg\{\int_0^T\int_0^\infty\l[1-2e^{bx}+e^{\gamma x}\r]\nu(dx)dt
      +a^2\sigma_0^2\calB(T)\bigg\}Y^{2a,\gamma}_T.
\end{align*}
Under (\ref{eq-lem4-2}), we have
$\int_1^\infty\exp\{2\gamma x\}\nu(dx)<\infty$.
Thus, we can see that $Y^{2a,\gamma}$ is a martingale
by the same sort argument as item 1.
Moreover, we have $\int_0^\infty\l[1-2e^{bx}+e^{\gamma x}\r]\nu(dx)<\infty$,
from which the square integrability of $Y^{a,b}_T$ follows.
\fin

Next, we give a definition of LRM based on Theorem 1.6 of Schweizer
\cite{Sch3}.

\begin{defn}
\begin{enumerate}
\item 
$\Theta_S$ denotes the space of all $\bbR$-valued predictable processes $\xi$
satisfying
$\bbE\l[\int_0^T\xi_t^2d\langle M\rangle_t+(\int_0^T|\xi_tdA_t|)^2\r]<\infty$.
\item
An $L^2$-strategy is given by a pair $\vp=(\xi, \eta)$,
where $\xi\in\Theta_S$ and $\eta$ is an adapted process such that
$V(\vp):=\xi S+\eta$ is a right continuous process
with $\bbE[V_t^2(\vp)]<\infty$ for every $t\in[0,T]$.
Note that $\xi_t$ (resp. $\eta_t$) represents the amount of units of
the risky asset (resp. the risk-free asset) an investor holds at time $t$.
\item
For claim $F\in L^2(\bbP)$, the process $C^F(\vp)$ defined by
$C^F_t(\vp):=F1_{\{t=T\}}+V_t(\vp)-\int_0^t\xi_sdS_s$
is called the cost process of $\vp=(\xi, \eta)$ for $F$.
\item
An $L^2$-strategy $\vp$ is said local risk-minimization (LRM) for claim $F$ if
$V_T(\vp)=0$ and $C^F(\vp)$ is a martingale orthogonal to $M$,
that is, $[C^F(\vp),M]$ is a uniformly integrable martingale.
\item
An $F\in L^2(\bbP)$ admits a F\"ollmer-Schweizer decomposition
(FS decomposition, for short) if it can be described by
\begin{equation}
\label{eqFS}
F=F_0+\int_0^T\xi_t^FdS_t+L_T^F,
\end{equation}
where $F_0\in\bbR$, $\xi^F\in\Theta_S$ and $L^F$ is a square-integrable
martingale orthogonal to $M$ with $L_0^F=0$.
\end{enumerate}
\end{defn}

\noindent
For more details on LRM, see Schweizer \cite{Sch}, \cite{Sch3}.
Now, we introduce a relationship between LRM and FS decomposition.

\begin{prop}
Under Assumption \ref{ass1}, LRM $\vp=(\xi,\eta)$ for $F$ exists
if and only if $F$ admits an FS decomposition; and its relationship is given by
\[
\xi_t=\xi^F_t,
\hspace{3mm}\eta_t=F_0+\int_0^t\xi^F_sdS_s+L^F_t-F1_{\{t=T\}}-\xi^F_tS_t.
\]
\end{prop}

\proof
This is from Proposition 5.2 of \cite{Sch3},
together with Proposition \ref{prop-SC}.
\fin

\noindent
Thus, it suffices to get a representation of $\xi^F$ in (\ref{eqFS})
in order to obtain LRM for claim $F$.
Henceforth, we identify $\xi^F$ with LRM for $F$.

\subsection{Minimal martingale measure}
We need to study upon the MMM in order to discuss FS decomposition.
A probability measure $\tP\sim\bbP$ is called the minimal martingale measure
(MMM), if $S$ is a $\tP$-martingale; and
any square-integrable $\bbP$-martingale orthogonal to $M$ remains
a martingale under $\tP$.
Now, we consider the following SDE:
\begin{equation}
\label{SDE-Z}
dZ_t=-Z_{t-}\Lambda_tdM_t, \ \ \ Z_0=1.
\end{equation}
The solution to (\ref{SDE-Z}) is a stochastic exponential of
$-\int_0^\cdot\Lambda_tdM_t$.
More precisely, denoting
\begin{equation}
\label{eq-ut}
u_t:=\Lambda_tS_{t-}\sigma_t
=\frac{\mu}{\sigma_t}+\l(\beta+\frac{1}{2}\r)\sigma_t
\end{equation}
for $t\in[0,T]$, we have $\Lambda_tdM_t=u_tdW_t$; and
\begin{equation}
\label{eq-ZT}
Z_t= \exp\bigg\{-\frac{1}{2}\int_0^tu_s^2ds-\int_0^tu_sdW_s\bigg\}.
\end{equation}
To see that $Z_T$ becomes the density of the MMM, it is enough to show
the square integrability of $Z_T$.

\begin{prop}
\label{prop-MMM}
$Z_T\in L^2(\bbP)$.
\end{prop}

\proof
First of all, there is a constant $C_u>0$ such that
\[
u^2_t=    \frac{\mu^2}{\sigma^2_t}+2\mu\l(\beta+\frac{1}{2}\r)
          +\l(\beta+\frac{1}{2}\r)^2\sigma_t^2
     \leq C_u+\l(\beta+\frac{1}{2}\r)^2\sigma_t^2
\]
by (\ref{eq-ut}).
Thus, (\ref{eq-ZT}) implies
\begin{align*}
Z^2_T
&=    \exp\l\{-2\int_0^Tu_t^2dt-\int_0^T2u_tdW_t+\int_0^Tu_t^2dt\r\}
      \\
&\leq \exp\l\{-2\int_0^Tu_t^2dt-\int_0^T2u_tdW_t
      +TC_u+\l(\beta+\frac{1}{2}\r)^2\int_0^T\sigma_t^2dt\r\} \\
&\leq \exp\l\{-2\int_0^Tu_t^2dt-\int_0^T2u_tdW_t
      +TC_u+\l(\beta+\frac{1}{2}\r)^2[\sigma_0^2\calB(T)+\calB(T)J_T]\r\} \\
&\leq \exp\bigg\{TC_u+\l(\beta+\frac{1}{2}\r)^2\sigma_0^2\calB(T)
      +\int_0^T\int_0^\infty\l[e^{\kappa x}-1\r]\nu(dx)dt\bigg\} \\
&     \hspace{7mm}\times\exp\bigg\{-2\int_0^Tu_t^2dt
      -\int_0^T2u_tdW_t+\int_0^T\int_0^\infty\kappa x\tN(dx,dt) \\
&     \hspace{7mm}+\int_0^T\int_0^\infty\l[\kappa x+1-e^{\kappa x}\r]
      \nu(dx)dt\bigg\},
\end{align*}
since $\l(\beta+\frac{1}{2}\r)^2\calB(T)\leq\kappa$ by (A2).
In addition, Remark \ref{rem1} implies
\begin{align*}
\bbE\l[\exp\l\{2\int_0^Tu^2_tdt\r\}\r]
&\leq \bbE\l[\exp\l\{2TC_u
      +2\l(\beta+\frac{1}{2}\r)^2\int_0^T\sigma^2_tdt\r\}\r] \\
&\leq \exp\l\{2TC_u+2\kappa\sigma_0^2\r\}\bbE\l[e^{2\kappa J_T}\r]
      <\infty.
\end{align*}
Hence, we can see that $Z_T^2$ is integrable
by the same manner as the proof of item 1 in Lemma \ref{lem4}.
\fin

Henceforth, we denote the MMM by $\tP$, that is,
we have $Z_T=\frac{d\tP}{d\bbP}$.
Note that $dW^{\tP}_t:=dW_t+u_tdt$ is a Brownian motion under $\tP$; and
$\tN$ remains a martingale under $\tP$.
Remark that we can rewrite (\ref{SDE}) and $L_T$ as
$dS_t=S_{t-}\sigma_tdW^{\tP}_t$ and
$L_T=\int_0^T\sigma_sdW^{\tP}_s-\frac{1}{2}\int_0^T\sigma^2_sds$, respectively.
The following lemma is indispensable to formulate a Malliavin calculus
under $\tP$.

\begin{lem}
\label{lem7}
$W^{\tP}$ is independent of $\tN$; and
$W^{\tP}_t+\int_0^t\int_0^\infty z\tN(ds,dz)(=:X^*_t)$ is a L\'evy process
under $\tP$.
\end{lem}

\proof
This is given from Theorem 3.2 in \cite{NV}.
Remark that Assumptions Z1 - Z3 in \cite{NV} are their standing assumptions.
Assumptions Z1 and Z2 are satisfied in our setting from Assumption \ref{ass1}.
On the other hand, Assumption Z3 does not necessarily hold,
but it is not needed for Theorem 3.2 in \cite{NV}.
\fin

\begin{rem}
The filtration $\bbF$ coincides with
the augmented filtration generated by $W^{\tP}$ and $\tN$.
\end{rem}

\subsection{Malliavin calculus under $\tP$}
Here, regarding $(\Omega,\calF,\tP)$ as the underlying probability space,
we formulate a Malliavin calculus for $X^*$ under $\tP$
based on Petrou \cite{Pet} and Chapter 5 of Renauld \cite{R}.
Although \cite{AS-BNS} adopted the canonical L\'evy space framework
undertaken by Sol\'e et al. \cite{S07},
we need to take a different way to define a Malliavin derivative,
since the property of the canonical L\'evy space is not preserved
under change of measure.

First of all, we need to prepare some notation; and define iterated integrals
with respect to  $W^{\tP}$ and $\tN$.
Denoting $U_0:=[0,T]$ and $U_1:=[0,T]\times(0,\infty)$, we define
\begin{align*}
Q_0(A)         &:= \int_AdW^{\tP}_t \hspace{3mm}\mbox{ for any }A\in\calB(U_0),
                                    \\
Q_1(A)         &:= \int_A\tN(dt,dx) \hspace{3mm}\mbox{ for any }A\in\calB(U_1),
                                    \\
\< Q_0\>       &:= m, \hspace{3mm}\mbox{ and }\hspace{3mm}
\< Q_1\>        := m\times \nu,
\end{align*}
where $m$ is the Lebesgue measure on $U_0$.
We denote
\[
G_{(j_1,\dots,j_n)}:=
\l\{(u^{j_1}_1,\dots,u^{j_n}_n)\in\prod_{k=1}^nU_{j_k}:0<t_1<\cdots<t_n<T\r\}
\]
for $n\in\bbN$ and $(j_1,\dots,j_n)\in\{0,1\}^n$,
where $u^{j_k}_k:=t_k$ if $j_k=0$; and $:=(t_k,x)$ if $j_k=1$
for $k=1,\dots,n$.
We define an $n$-fold iterated integral as follows:
\[
J_n^{(j_1,\dots,j_n)}(g_n^{(j_1,\dots,j_n)})
:=\int_{G_{(j_1,\dots,j_n)}}g_n^{(j_1,\dots,j_n)}(u^{j_1}_1,\dots,u^{j_n}_n)
  Q_{j_1}(du_1^{j_1})\cdots Q_{j_n}(du_n^{j_n}),
\]
where $g_n^{(j_1,\dots,j_n)}$ is a deterministic function
in $L^2\l(G_{(j_1,\dots,j_n)},\bigotimes_{k=1}^n\< Q_{j_k}\> \r)$.
Then, Theorem 1 in \cite{Pet} ensures that
every $L^2(\tP)$ random variable $F$ is represented as a sum of
iterated integrals, that is, we can find deterministic functions
$g_n^{(j_1,\dots,j_n)}\in
L^2\l(G_{(j_1,\dots,j_n)},\bigotimes_{k=1}^n\< Q_{j_k}\> \r)$
for $n\in\bbN$ and $(j_1,\dots,j_n)\in\{0,1\}^n$ such that
$F$ has the following chaos expansion:
\begin{equation}
\label{chaos}
F=\bbE_{\tP}[F]+\sum_{n=1}^\infty\sum_{(j_1,\dots,j_n)\in\{0,1\}^n}
  J_n^{(j_1,\dots,j_n)}(g_n^{(j_1,\dots,j_n)}).
\end{equation}
Note that the infinite series in (\ref{chaos}) converges in $L^2(\tP)$.

Now, we define $\bbD^0$ the space of Malliavin differentiable random variables;
and a Malliavin derivative operator $D^0$.
Denoting, for $1\leq k\leq n$ and $t\in(0,T)$,
\begin{align*}
G^k_{(j_1,\dots,j_n)}(t)
&:= \big\{(u^{j_1}_1,\dots,u^{j_{k-1}}_{k-1},u^{j_{k+1}}_{k+1}\dots,u^{j_n}_n)
    \in G_{(j_1,\dots,j_{k-1},j_{k+1},\dots,j_n)}: \\
&   \hspace{7mm}0<t_1<\cdots<t_{k-1}<t<t_{k+1}<\cdots<t_n<T\big\},
\end{align*}
we define $\bbD^0$ as
\begin{align*}
\bbD^0
&:= \bigg\{F\in L^2(\tP),
    F=\bbE_{\tP}[F]+\sum_{n=1}^\infty\sum_{(j_1,\dots,j_n)\in\{0,1\}^n}
    J_n^{(j_1,\dots,j_n)}(g_n^{(j_1,\dots,j_n)}): \\
&   \hspace{7mm}\|g_1^{(0)}\|^2_{L^2(m)}+\sum_{n=2}^\infty
    \sum_{(j_1,\dots,j_n)\in\{0,1\}^n}\sum_{k=1}^n{\bf 1}_{\{j_k=0\}} \\
&   \hspace{7mm}\times\int_0^T\l\|g_n^{(j_1,\dots,j_{k-1},0,j_{k+1},\dots,j_n)}
    (\dots,t,\dots)\r\|^2_{L^2\l(G^k_{(j_1,\dots,j_n)}(t)\r)}dt
    <\infty\bigg\}.
\end{align*}
Moreover, for $F\in\bbD^0$ and $t\in[0,T]$, we define
\begin{align*}
D^0_tF
&:= g_1^{(0)}(t)
    +\sum_{n=2}^\infty\sum_{(j_1,\dots,j_n)\in\{0,1\}^n}\sum_{k=1}^n
    {\bf 1}_{\{j_k=0\}} \\
&   \hspace{7mm}\times J_{n-1}^{(j_1,\dots,j_{k-1},j_{k+1},\dots,j_n)}
    \l(g_n^{(j_1,\dots,j_{k-1},0,j_{k+1},\dots,j_n)}(\dots,t,\dots)
    {\bf 1}_{G^k_{(j_1,\dots,j_n)}(t)}\r).
\end{align*}

\setcounter{equation}{0}
\section{Main results}
We give explicit representations of LRM for call and put options
as our main results.
As in \cite{AS-BNS}, we consider firstly put options,
since a Malliavin derivative for put options is given
owing to its boundedness.
LRM for call options will be given as a corollary.
If we dealt with call options firstly, then we would need to impose
additional assumptions.

Before stating our main theorem, we prepare two propositions,
one is a Malliavin derivative for put options; and
the other is a Clark-Ocone type representation result for random variables
in $\bbD^0$.

\begin{prop}
\label{prop-put}
For $K>0$, we have $(K-S_T)^+\in\bbD^0$, and
\[
D^0_t(K-S_T)^+= -{\bf 1}_{\{S_T<K\}}S_T\sigma_t.
\]
\end{prop}

\proof
The same result has been given in Proposition 4.1 of \cite{AS-BNS}.
However, their framework of Malliavin calculus is different from ours
as said at the beginning of Subsection 2.3.
Thus, we give a proof anew by the same way as \cite{AS-BNS}.

First of all, by the same argument as Lemma A.1 in \cite{AS-BNS},
we have $D^0_t\sigma^2_s=0$.
Theorem 2 in \cite{Pet} implies $D^0_t\sigma_s=0$
by the same manner as Lemma A.2 of \cite{AS-BNS}.
In addition, by the same way as Lemmas A.3 and A.4 in \cite{AS-BNS},
we can see that $D^0_t\int_0^T\sigma_s^2ds=0$;
and $D^0_t\int_0^T\sigma_sdW^{\tP}_s=\sigma_t$
by using Proposition 6 in \cite{Pet}.
As a result, we obtain $L_T\in\bbD^0$ and $D^0_tL_T=\sigma_t$.
Next, denoting
\[
f_K(r):=
\begin{cases}
S_0e^r,              & \mbox{if } r\leq\log(K/S_0), \\
Kr+K(1-\log(K/S_0)), & \mbox{if } r>\log(K/S_0).
\end{cases}
\] 
we have that $f_K\in C^1(\bbR)$ and $0<f^\prime_K(r)\leq K$ for any $r\in\bbR$.
Thus, Theorem 2 of \cite{Pet} implies that
$f_K(L_T)\in\bbD^0$ and
\begin{equation}
\label{eq-lem2-1}
D^0_tf_K(L_T)=f^\prime_K(L_T)D^0_tL_T=f^\prime_K(L_T)\sigma_t.
\end{equation}

Since $(K-S_T)^+=(K-f_K(L_T))^+$, we need only to see
$(K-f_K(L_T))^+\in\bbD^0$; and calculate $D^0_t(K-f_K(L_T))^+$.
To this end, we take a mollifier function $\vp$ which
is a $C^\infty$-function from $\bbR$ to $[0,\infty)$ with
$\supp(\vp)\subset[-1,1]$ and $\int_{-\infty}^\infty\vp(x)dx=1$.
We denote $\vp_n(x):=n\vp(nx)$ and
$g_n(x):=\int_{-\infty}^\infty (K-y)^+\vp_n(x-y)dy$ for any $n\geq1$.
Noting that
\[
g_n(x)=\int_{-\infty}^\infty\l(K-x+\frac{y}{n}\r)^+\vp(y)dy
      = \int_{-n(K-x)}^\infty\l(K-x+\frac{y}{n}\r)\vp(y)dy,
\]
we have $g^\prime_n(x)=-\int_{-n(K-x)}^\infty\vp(y)dy$,
so that $g_n\in C^1$ and $|g^\prime_n|\leq1$.
Thus, Theorem 2 in \cite{Pet} again implies that,
for any $n\geq1$, $g_n(f_K(L_T))\in\bbD^0$ and
\begin{equation}
\label{eq-lem2-2}
D^0_tg_n(f_K(L_T))=g^\prime_n(f_K(L_T))D^0_tf_K(L_T)
                  =g^\prime_n(f_K(L_T))f^\prime_K(L_T)\sigma_t
\end{equation}
by (\ref{eq-lem2-1}).
We have then
\[
\sup_{n\geq1}\|D^0g_n(f_K(L_T))\|_{L^2(m\times\tP)}^2
\leq K^2\bbE_{\tP}\l[\int_0^T\sigma_t^2dt\r]<\infty.
\]
In addition, noting that
\begin{align*}
|g_n(x)-(K-x)^+|
&=    \l|\int_{-1}^1\l\{\l(K-x+\frac{y}{n}\r)^+-(K-x)^+\r\}\vp(y)dy\r| \\
&\leq \frac{1}{n}\int_{-1}^1|y|\vp(y)dy\leq\frac{1}{n}
\end{align*}
for any $x\in\bbR$,
we have $\lim_{n\to\infty}\bbE[|g_n(f_K(L_T))-(K-f_K(L_T))^+|^2]=0$.
As a result, Lemma \ref{lem5} below implies that $(K-f_K(L_T))^+\in\bbD^0$.
Furthermore, Lemma 2 of \cite{Pet} ensures the existence of a subsequence $n_k$
such that $D^0g_{n_k}(f_K(L_T))$ converges to $D^0(K-f_K(L_T))^+$
in the sense of $L^2(m\times\tP)$.
On the other hand, we have $\lim_{n\to\infty}g^\prime_n(x)
=-{\bf 1}_{\{x<K\}}-{\bf 1}_{\{x=K\}}\int^\infty_0\vp(y)dy$;
and $\tP(f_K(L_T)=K)=0$ by Corollary 2.3 of \cite{NV},
from which $\lim_{n\to\infty}g^\prime_n(f_K(L_T))=-{\bf 1}_{\{f_K(L_T)<K\}}$
a.s. follows.
Consequently, by taking a further subsequence if need be,
(\ref{eq-lem2-2}) provides
\begin{align*}
D^0_t(K-S_T)^+
&= D^0_t(K-f_K(L_T))^+
   = \lim_{k\to\infty}D^0_tg_{n_k}(f_K(L_T)) \\
&= \lim_{k\to\infty}g^\prime_{n_k}(f_K(L_T))f^\prime_K(L_T)\sigma_t
   = -{\bf 1}_{\{f_K(L_T)<K\}}f^\prime_K(L_T)\sigma_t \\
&= -{\bf 1}_{\{S_T<K\}}S_T\sigma_t,
   \hspace{7mm}m\times\tP\mbox{-a.s.}
\end{align*}
\fin

\begin{lem}
\label{lem5}
Let $F$ be in $L^2(\tP)$, and $(F_n)_{n\geq1}$ a sequence of $\bbD^0$
converging to $F$ in $L^2(\tP)$.
If $\sup_{n\geq1}\|D^0F_n\|_{L^2(m\times\tP)}<\infty$,
then $F\in\bbD^0$.
\end{lem}

\proof
This is given from the proof of Lemma 5.5.5 of \cite{R}.
\fin

\begin{prop}
\label{prop-CO}
For $F\in\bbD^0$, we have
\[
F=\bbE_{\tP}[F]+\int_0^T\bbE_{\tP}[D^0_tF|\calF_{t-}]dW^{\tP}_t
  +\int_0^T\int_0^\infty\psi_{t,x}\tN(dt,dx)
\]
for some predictable process $\psi\in L^2(m\times\nu\times\tP)$.
\end{prop}

\proof
Denoting by (\ref{chaos}) the chaos expansion of $F$, we have
\begin{align}
\label{eq-prop-CO}
\footnotesize{\mbox{$F$}}
&\footnotesize{\mbox{$=$}}\footnotesize{\mbox{$\displaystyle{
     \bbE_{\tP}[F]+\sum_{n=1}^\infty\sum_{(j_1,\dots,j_{n-1})\in\{0,1\}^{n-1}}
     \l\{J_n^{(j_1,\dots,j_{n-1},0)}(g_n^{(j_1,\dots,j_{n-1},0)})
     +J_n^{(j_1,\dots,j_{n-1},1)}(g_n^{(j_1,\dots,j_{n-1},1)})\r\}}$}}
     \nonumber \\
&\footnotesize{\mbox{$=$}}\footnotesize{\mbox{$\displaystyle{
    \bbE_{\tP}[F]+\int_0^Tg_1^{(0)}(t)dW^{\tP}_t+\sum_{n=2}^\infty
    \sum_{(j_1,\dots,j_{n-1})\in\{0,1\}^{n-1}}\int_0^T
    J_{n-1}^{(j_1,\dots,j_{n-1})}\l(g_n^{(j_1,\dots,j_{n-1},0)}(\dots,t)
    {\bf 1}_{G^n_{(j_1,\dots,j_n)}(t)}\r)dW^{\tP}_t}$}} \nonumber \\
&   \hspace{3mm}\footnotesize{\mbox{$\displaystyle{
    +\int_0^T\int_0^\infty g_1^{(1)}((t,x))\tN(dt,dx)}$}} \nonumber \\
&   \hspace{3mm}\footnotesize{\mbox{$\displaystyle{
    +\sum_{n=2}^\infty\sum_{(j_1,\dots,j_{n-1})\in\{0,1\}^{n-1}}
    \int_0^T\int_0^\infty J_{n-1}^{(j_1,\dots,j_{n-1})}
    \l(g_n^{(j_1,\dots,j_{n-1},1)}(\dots,(t,x))
    {\bf 1}_{G^n_{(j_1,\dots,j_n)}(t)}\r)\tN(dt,dx)}$}} \nonumber \\
&\footnotesize{\mbox{$=$}}\footnotesize{\mbox{$\displaystyle{
    \bbE_{\tP}[F]+\int_0^T\bigg\{g_1^{(0)}(t)+\sum_{n=2}^\infty
    \sum_{(j_1,\dots,j_{n-1})\in\{0,1\}^{n-1}}
    J_{n-1}^{(j_1,\dots,j_{n-1})}\l(g_n^{(j_1,\dots,j_{n-1},0)}(\dots,t)
    {\bf 1}_{G^n_{(j_1,\dots,j_n)}(t)}\r)\bigg\}dW^{\tP}_t}$}} \nonumber \\
&   \hspace{3mm}\footnotesize{\mbox{$\displaystyle{
    +\int_0^T\int_0^\infty\bigg\{g_1^{(1)}((t,x))
    +\sum_{n=2}^\infty\sum_{(j_1,\dots,j_{n-1})\in\{0,1\}^{n-1}}
    J_{n-1}^{(j_1,\dots,j_{n-1})}\l(g_n^{(j_1,\dots,j_{n-1},1)}(\dots,(t,x))
    {\bf 1}_{G^n_{(j_1,\dots,j_n)}(t)}\r)\bigg\}\tN(dt,dx)}$}} \\
&\footnotesize{\mbox{$=:$}}\footnotesize{\mbox{$\displaystyle{
    \bbE_{\tP}[F]+\int_0^T\phi_tdW^{\tP}_t
    +\int_0^T\int_0^\infty\psi_{t,x}\tN(dt,dx)}$}}. \nonumber
\end{align}
The above third equality (\ref{eq-prop-CO}) is proved
in Lemma \ref{lem6} below.
On the other hand, noting that $F\in\bbD^0$, we have
{\scriptsize\begin{align*}
\footnotesize{\mbox{$\bbE_{\tP}[D^0_tF|\calF_{t-}]$}}
&\footnotesize{\mbox{$=$}}\footnotesize{\mbox{$\displaystyle{
   \bbE_{\tP}\bigg[g_1^{(0)}(t)
   +\sum_{n=2}^\infty\sum_{(j_1,\dots,j_n)\in\{0,1\}^n}\sum_{k=1}^n
   {\bf 1}_{\{j_k=0\}}}$}} \\
&  \hspace{3mm}\footnotesize{\mbox{$\displaystyle{
   \times J_{n-1}^{(j_1,\dots,j_{k-1},j_{k+1},\dots,j_n)}
   \bigg(g_n^{(j_1,\dots,j_{k-1},0,j_{k+1},\dots,j_n)}(\dots,t,\dots)
   {\bf 1}_{G^k_{(j_1,\dots,j_n)}(t)}\bigg)\bigg|\calF_{t-}\bigg]}$}} \\
&\footnotesize{\mbox{$=$}}\footnotesize{\mbox{$\displaystyle{
   g_1^{(0)}(t)+\sum_{n=2}^\infty\sum_{(j_1,\dots,j_n)\in\{0,1\}^n}\sum_{k=1}^n
   {\bf 1}_{\{j_k=0\}}}$}} \\
&  \hspace{3mm}\footnotesize{\mbox{$\displaystyle{
   \times \bbE_{\tP}\bigg[J_{n-1}^{(j_1,\dots,j_{k-1},j_{k+1},\dots,j_n)}
   \bigg(g_n^{(j_1,\dots,j_{k-1},0,j_{k+1},\dots,j_n)}(\dots,t,\dots)
   {\bf 1}_{G^k_{(j_1,\dots,j_n)}(t)}\bigg)\bigg|\calF_{t-}\bigg]}$}} \\
&\footnotesize{\mbox{$=$}}\footnotesize{\mbox{$\displaystyle{
   g_1^{(0)}(t)+\sum_{n=2}^\infty\sum_{(j_1,\dots,j_n)\in\{0,1\}^n}
   {\bf 1}_{\{j_n=0\}}J_{n-1}^{(j_1,\dots,j_{n-1})}
   \bigg(g_n^{(j_1,\dots,j_{n-1},0)}(\dots,t)
   {\bf 1}_{G^n_{(j_1,\dots,j_n)}(t)}\bigg)}$}} \\
&\footnotesize{\mbox{$=$}}\footnotesize{\mbox{$
   \phi_t$}}.
\end{align*}}
As a result, $\phi$ belongs to $L^2(m\times\tP)$.
Thus, $\int_0^T\int_0^\infty\psi_{t,x}\tN(dt,dx)$ is square integrable,
that is, $\psi\in L^2(m\times\nu\times\tP)$.
This completes the proof of Proposition \ref{prop-CO}.
\fin

\begin{lem}
\label{lem6}
(\ref{eq-prop-CO}) in the proof of Proposition \ref{eq-prop-CO} holds true.
In other words, we have, for $l=0,1$,
\begin{align*}
\lefteqn{\sum_{n=2}^\infty\sum_{(j_1,\dots,j_{n-1})\in\{0,1\}^{n-1}}\int_{U_l}
J_{n-1}^{(j_1,\dots,j_{n-1})}\l(g_n^{(j_1,\dots,j_{n-1},l)}
(\dots,\widehat{u}^l){\bf 1}_{G^n_{(j_1,\dots,j_n)}(t)}\r)Q_l(d\widehat{u}^l)}
\\
&=
\int_{U_l}\sum_{n=2}^\infty\sum_{(j_1,\dots,j_{n-1})\in\{0,1\}^{n-1}}
J_{n-1}^{(j_1,\dots,j_{n-1})}\l(g_n^{(j_1,\dots,j_{n-1},l)}
(\dots,\widehat{u}^l){\bf 1}_{G^n_{(j_1,\dots,j_n)}(t)}\r)Q_l(d\widehat{u}^l),
\end{align*}
where $\widehat{u}^0=t\in U_0$ and $\widehat{u}^1=(t,x)\in U_1$.
\end{lem}

\proof
Recall that the infinite series in a chaos expansion converges
in the $L^2(\tP)$-sense.
Now, for $l=0,1$, we denote
\[
\Phi^{l,N}(\widehat{u}^l)
:=\sum_{n=2}^N\sum_{(j_1,\dots,j_{n-1})\in\{0,1\}^{n-1}}
J_{n-1}^{(j_1,\dots,j_{n-1})}\l(g_n^{(j_1,\dots,j_{n-1},l)}
(\dots,\widehat{u}^l){\bf 1}_{G^n_{(j_1,\dots,j_n)}(t)}\r)
\]
for $N\geq2$, and
\[
\Phi^l(\widehat{u}^l)
:=\sum_{n=2}^\infty\sum_{(j_1,\dots,j_{n-1})\in\{0,1\}^{n-1}}
J_{n-1}^{(j_1,\dots,j_{n-1})}\l(g_n^{(j_1,\dots,j_{n-1},l)}
(\dots,\widehat{u}^l){\bf 1}_{G^n_{(j_1,\dots,j_n)}(t)}\r).
\]
We have then that, for $l=0,1$, $(\Phi^{l,N})_{N\geq2}$ is a sequence of
$L^2(\< Q_l\> \times\tP)$ converging to $\Phi^l$
in the $L^2(\< Q_l\> \times\tP)$-sense.
Thus, we have
\[
\lim_{N\to\infty}
\bbE_{\tP}\l[\l|\int_{U_l}\Phi^{l,N}(\widehat{u}^l)Q_l(d\widehat{u}^l)
-\int_{U_l}\Phi^l(\widehat{u}^l)Q_l(d\widehat{u}^l)\r|^2\r]=0.
\]
\fin

The following theorem is our main result.

\begin{thm}
\label{thm-main}
For $K>0$, LRM $\xi^{(K-S_T)^+}$ of put option $(K-S_T)^+$ is represented as
\begin{equation}
\label{eq-thm-main}
\xi_t^{(K-S_T)^+}
=\frac{-1}{S_{t-}}\bbE_{\bbP^*}[{\bf 1}_{\{S_T<K\}}S_T|\calF_{t-}].
\end{equation}
\end{thm}

\proof
Denoting by $\zeta_t$ the right hand side of (\ref{eq-thm-main}),
we shall see that the process $\zeta$ is in $\Theta_S$.
Noting that $|\zeta_t|\leq\frac{K}{S_{t-}}$, we have
\begin{align*}
\lefteqn{\bbE\l[\int_0^T\zeta_t^2d\langle M\rangle_t
+\l(\int_0^T|\zeta_tdA_t|\r)^2\r]} \\
&\leq \bbE\l[\int_0^TK^2\sigma^2_tdt
      +\l(\int_0^TK\l|\mu+\l(\beta+\frac{1}{2}\r)\sigma^2_t\r|dt\r)^2\r]
      <\infty,
\end{align*}
since $\bbE\l[\l(\int_0^T\sigma^2_tdt\r)^2\r]<\infty$ by Lemma \ref{lem3}.
As a result, $\zeta\in \Theta_S$ holds.

Next, defining
\[
L^{(K-S_T)^+}_t:=\bbE\l[(K-S_T)^+
                 -\bbE_{\tP}\l[(K-S_T)^+\r]-\int_0^T\zeta_sdS_s\Big|\calF_t\r],
\]
we show that
\[
(K-S_T)^+=\bbE_{\tP}\l[(K-S_T)^+\r]+\int_0^T\zeta_tdS_t+L^{(K-S_T)^+}_T
\]
gives an FS decomposition of $(K-S_T)^+$.
Since $L^{(K-S_T)^+}$ is a $\bbP$-martingale with
$L^{(K-S_T)^+}_T\in L^2(\bbP)$,
we have only to show the orthogonality of $L^{(K-S_T)^+}$ to $M$.
Since $(K-S_T)^+\in\bbD^0$ from Proposition \ref{prop-put},
we have, by Propositions \ref{prop-CO} and \ref{prop-put},
\begin{align*}
(K-S_T)^+
&= \bbE_{\tP}\l[(K-S_T)^+\r]
   +\int_0^T\bbE_{\tP}\l[D^0_t(K-S_T)^+|\calF_{t-}\r]dW^{\tP}_t \\
&  \hspace{5mm}+\int_0^T\int_0^\infty\psi_{t,x}\tN(dt,dx) \\
&= \bbE_{\tP}\l[(K-S_T)^+\r]
   -\int_0^T\bbE_{\tP}\l[{\bf 1}_{\{S_T<K\}}S_T|\calF_{t-}\r]\sigma_tdW^{\tP}_t
   \\
&  \hspace{5mm}+\int_0^T\int_0^\infty\psi_{t,x}\tN(dt,dx) \\
&= \bbE_{\tP}\l[(K-S_T)^+\r]+\int_0^T\zeta_tdS_t
   +\int_0^T\int_0^\infty\psi_{t,x}\tN(dt,dx)
\end{align*}
for some predictable process $\psi\in L^2(m\times\nu\times\tP)$,
which means $L^{(K-S_T)^+}_t=\int_0^t\int_0^\infty\psi_{s,x}\tN(ds,dx)$
for any $t\in[0,T]$.
Thus, $L^{(K-S_T)^+}$ is orthogonal to $M$.
\fin

\noindent
By the put-call parity, the following holds:

\begin{cor}
\label{cor1}
LRM for call option $(S_T-K)^+$ is given as
$\xi^{(S_T-K)^+}=1+\xi^{(K-S_T)^+}$.
\end{cor}

\setcounter{equation}{0}
\section{Conclusions}
We give representations of LRM of call and put options
for BNS models with constraint $\rho=0$.
Compared with \cite{AS-BNS}, we relax the restriction on $\beta$; and 
restrict $\rho$ to $0$ instead.
The representation (\ref{eq-thm-main}) in Theorem \ref{thm-main} coincides with
(3.1) in Theorem 3.1 of \cite{AS-BNS} by substituting $0$ for $\rho$.
Note that $\beta$ does not appear in representations of LRM,
although the density of the MMM is depending on $\beta$.

Some important problems related to LRM for BNS models still remains
to future research:
development of numerical scheme, comparison with delta hedge,
extensions to the fully general case of BNS models, and so forth.

\begin{center}
{\bf Acknowledgements}
\end{center}
The author would like to thank to Jean-Pierre Fouque for fruitful discussion;
and acknowledge the financial support by
Ishii memorial securities research promotion foundation.



\begin{thebibliography}{00}
\bibitem{AS-BNS}
Arai, T., Suzuki, R.:
Local risk minimization for Barndorff-Nielsen and Shephard models.
submitted. Available at
\url{http://arxiv.org/pdf/1503.08589v1}
\bibitem{BNS1}
Barndorff-Nielsen, O.E., Shephard, N.:
Modelling by L\'evy processes for financial econometrics. 
In: Barndorff-Nielsen, O.E., Mikosch,T., Resnick, S. (eds.):
L\'evy processes --Theory and Applications, pp. 283--318.
Birkh\"auser, Basel (2001)
\bibitem{BNS2}
Barndorff-Nielsen, O.E., Shephard, N.:
Non-Gaussian Ornstein-Uhlenbeck based models and some of their uses
in financial econometrics. J.R. Statistic. Soc. 63, 167--241 (2001)
\bibitem{CT}
Cont R., Tankov P.,
Financial Modelling with Jump Processes.
Chapman \& Hall, London (2004)
\bibitem{I}
Ishikawa, Y.: Stochastic Calculus of Variations for Jump Processes.
Walter De Gruyter, Berlin (2013)
\bibitem{NV}
Nicolato, E., Venardos, E.: 
Option Pricing in Stochastic Volatility Models of the Ornstein-Uhlenbeck type.
Math. Finance. 13 (4), 445--466 (2003)
\bibitem{Pet}
Petrou, E.
Malliavin calculus in L\'evy spaces and applications to finance.
Electronic Journal of Probability. 27, 852-879 (2008)
\bibitem{R}
Renaud, J.F.:
Calcul de Malliavin, processus de L\'evy et applications en finance:
quelques contributions.
Dissertation, Universit\'e de Montr\'eal (2007)
Available at
\url{http://neumann.hec.ca/pages/bruno.remillard/Theses/JFRenaud.pdf}
\bibitem{Scho}
Schoutens, W.:
L\'evy Processes in Finance: Pricing Financial Derivatives.
John Wiley \& Sons, Hoboken (2003)
\bibitem{Sch}
Schweizer, M.:
A Guided Tour through Quadratic Hedging Approaches.
In: Jouini, E., Cvitani{\'c}, J., Musiela, M. (eds.):
Option Pricing, Interest Rates
and Risk Management (Handbooks in Mathematical Finance), pp. 538--574.
Cambridge University Press, Cambridge (2001)
\bibitem{Sch3}
Schweizer, M.:
Local Risk-Minimization for Multidimensional Assets and Payment Streams.
Banach Center Publ. 83, 213--229 (2008)
\bibitem{S07}
Sol\'e, J.L., Utzet, F., Vives, J.: 
Canonical L\'evy process and Malliavin calculus. 
Stochastic Process. Appl. 117, 165--187 (2007)
\bibitem{WQW}
Wang, W., Qian,L. and Wang, W.:
Hedging strategy for unit-linked life insurance contracts in stochastic
volatility models.
WSEAS Transactions on Mathematic, Volume 12, Issue 4, (2013)
\end{thebibliography}
\end{document}